\documentclass[aip,jap,amsmath,amssymb,reprint,a4paper,noshowpacs,citeautoscript]{revtex4-1}
\usepackage[sc]{mathpazo}
\usepackage[scaled=0.9]{helvet}
\usepackage[utf8]{inputenc}
\usepackage{graphicx}
\usepackage{xspace}
\usepackage[
,textwidth=17.5cm
,textheight=23.5cm
,verbose
,dvips
]{geometry}

\usepackage{textcomp}

\newcommand{\ingax}{In$_{x}$Ga$_{1-x}$N\xspace}
\newcommand{\ingay}{In$_{y}$Ga$_{1-y}$N\xspace}
\newcommand{\ingaxy}{In$_{x}$Ga$_{1-x}$N/In$_{y}$Ga$_{1-y}$N\xspace}
\newcommand{\ingaxyo}{In$_{x}$Ga$_{1-x}$N/In$_{y}$Ga$_{1-y}$N$(000\bar{1})$\xspace}
\newcommand{\ingayxo}{In$_{y}$Ga$_{1-y}$N/In$_{x}$Ga$_{1-x}$N$(000\bar{1})$\xspace}

\begin{document}

\title{Comparison of the luminous efficiency of Ga- and N-polar \ingaxy quantum wells grown by plasma-assisted molecular beam epitaxy}

\author{Sergio Fernández-Garrido}
\email{garrido@pdi-berlin.de}
\author{Jonas Lähnemann}
\altaffiliation[Present address: ]{Equipe mixte CEA-CNRS-UJF Nanophysique et Semiconducteurs, INAC/SP2M, CEA-Grenoble, 17 rue des Martyrs, 38054 Grenoble, France}
\author{Christian Hauswald}
\affiliation{Paul-Drude-Institut für Festkörperelektronik,
Hausvogteiplatz 5--7, 10117 Berlin, Germany}
\author{Maxim Korytov}
\altaffiliation[Present address: ]{CEMES-CNRS and Université de Toulouse, 29 rue Jeanne Marvig, F-31055 Toulouse, France}
\author{Martin Albrecht}
\affiliation{Leibniz-Institut für Kristallzüchtung, Max-Born-Strasse 2, 12489 Berlin, Germany}
\author{Caroline Chèze}
\altaffiliation[Present address: ]{Paul-Drude-Institut für Festkörperelektronik, Hausvogteiplatz 5--7, 10117 Berlin, Germany}
\author{Czes{\l}aw Skierbiszewski}
\affiliation{TopGaN sp. z.o.o., ul. Soko{\l}owska 29/37, 01-142 Warszawa, Poland}
\author{Oliver Brandt}
\affiliation{Paul-Drude-Institut für Festkörperelektronik,
Hausvogteiplatz 5--7, 10117 Berlin, Germany}

\date{\today}

\begin{abstract}

We investigate the luminescence of Ga- and N-polar \ingaxy quantum wells grown by plasma-assisted molecular beam epitaxy on freestanding GaN as well as 6H-SiC substrates. In striking contrast to their Ga-polar counterparts, the N-polar quantum wells prepared on freestanding GaN do not exhibit any detectable photoluminescence even at 10~K. Theoretical simulations of the band profiles combined with resonant excitation of the quantum wells allow us to rule out carrier escape and subsequent surface recombination as the reason for this absence of luminescence. To explore the hypothesis of a high concentration of nonradiative defects at the interfaces between wells and barriers, we analyze the photoluminescence of Ga- and N-polar quantum wells prepared on 6H-SiC as a function of the well width. Intense luminescence is observed for both Ga- and N polar samples. As expected, the luminescence of the Ga-polar quantum wells quenches and red-shifts with increasing well width due to the quantum confined Stark effect. In contrast, both the intensity and the energy of the luminescence from the N-polar samples are essentially independent of well width. Transmission electron microscopy reveals that the N-polar quantum wells exhibit abrupt interfaces and homogeneous composition, excluding emission from In-rich clusters as the reason for this anomalous behavior. The microscopic origin of the luminescence in the N-polar samples is elucidated using spatially resolved cathodoluminescence spectroscopy.
Regardless of well width, the luminescence is found to not originate from the N-polar quantum wells, but from the semipolar facets of $\vee$-pit defects. These results cast serious doubts on the potential of N-polar \ingaxy quantum wells grown by plasma-assisted molecular beam epitaxy for the development of long-wavelength light emitting diodes. What remains to be seen is whether unconventional growth conditions may enable a significant reduction in the concentration of nonradiative defects.
\end{abstract}


\maketitle

\section{Introduction}

The growth of \ingax alloys along the $[000\bar{1}]$ direction (known as N-polar orientation) with high In contents has regained interest because it offers potential advantages for the fabrication of green light-emitting diodes (LEDs). Despite the rougher surfaces and higher concentrations of impurities in N-polar group-III nitride films as compared to their Ga-polar (i.\,e., $[0001]$ oriented) counterparts grown under identical conditions,\cite{Li2000,Ng2002,Masui_JJAPL_2009,Cheze_JVST_01_2013,Wong2013} the $[000\bar{1}]$ orientation offers two interesting advantages. The first of these advantages is an enhanced In-incorporation efficiency.\cite{Nath_01_APL_2010,Wong2013} This phenomenon is the result of the higher thermal stability of N-polar InN and paves the way for the use of significantly higher substrate temperatures\cite{Koblmueller_JAP_01_2007} that may help to improve the crystal quality as well as to minimize the incorporation of impurities. The second advantage is related to the expected improvement in device performance caused by the reversed direction of the polarization fields in the quantum wells (QWs) that constitute the active region of LEDs. As discussed in Ref.~\onlinecite{Han_01_JJAP_2012}, this fact should result in lower turn-on voltages and higher internal quantum efficiencies. Unlike other epitaxial growth techniques, where the incorporation of high In contents is challenging due to the low decomposition efficiency of NH$_{3}$ at reduced substrate temperatures,\cite{Dupuis_JCG_01_1997} plasma-assisted molecular beam epitaxy (PA-MBE) facilitates the synthesis of \ingax alloys across the entire compositional range.\cite{Iliopoulos_PSSa_01_2006,Aseev_01_APL_2015,Fabien_01_JCG_2015} 

Inspired by the potential advantages of N-polar \ingax alloys, \citet{Akyol_01_JJAP_2011} demonstrated the PA-MBE growth of N-polar \ingax LEDs emitting at $540$~nm in 2011.\cite{Akyol_01_JJAP_2011} However, in contrast to this encouraging work, some disconcerting results have been recently reported by \citet{Cheze_JVST_01_2013} who analyzed the optical properties of \ingaxy QWs grown by PA-MBE along the $[0001]$ and $[000\bar{1}]$ directions. Despite a comparable structural and morphological quality of their Ga- and N-polar QWs, the latter did not exhibit any detectable photoluminescence (PL) associated with the \ingaxy QWs. In contrast, an unclad \emph{thick} (150~nm) N-polar \ingax layer grown under identical conditions did show intense emission in the red spectral range. Therefore, \citet{Cheze_JVST_01_2013} attributed the lack of luminescence in the N-polar QWs to either surface-induced electric fields, causing carriers to escape from the QWs and to subsequently recombine at the surface nonradiatively, or to a high concentration of non-radiative defects located at the interfaces between wells and barriers. The poor luminous efficiency of N-polar \ingax QWs does not seem to be a specific problem of PA-MBE, but has also been reported for N-polar \ingaxy QWs grown by metal organic chemical vapor deposition (MOCVD).\cite{Keller_APL_2007,Masui_JJAPL_2009,Song_01_AMI_2015} Also for the case of MOCVD, this phenomenon is not understood yet, but has been tentatively attributed to higher dislocation densities\cite{Keller_APL_2007} as well as to elevated residual impurity concentrations.\cite{Keller_APL_2007,Masui_JJAPL_2009}

In this work, we investigate the luminescence of Ga- and N-polar \ingaxy QWs to find a consistent explanation for the differences in their efficiencies. Carrier escape and surface recombination is ruled out by combining simulations of the band profiles with PL measurements carried out with resonant excitation. To explore the remaining possibility, namely, the presence of a high concentration of nonradiative defects at the interfaces between wells and barriers, we analyze Ga- and N-polar QWs with different widths. The PL spectra of Ga-polar QWs exhibits the expected behavior, i.\,e., it quenches and red-shifts with increasing QW width. For the N-polar samples, we observe an intense emission whose intensity and energy do not depend on the QW width. The analysis of the N-polar QWs by transmission electron microscopy (TEM) reveal a random alloy composition and the absence of clustering. The anomalous behavior of the PL is thus not caused by the localization of excitons at alloy inhomogenities. The actual origin of the luminescence is clarified using spatially resolved cathodoluminescence (CL) spectroscopy. These measurements show that, regardless of the QW width, the luminescence does not arise from the N-polar QWs themselves but from semipolar ones formed in $\vee$-pit defects.

\section{Experiments and Methods}

\begin{table}
\caption{Substrates as well as widths \textit{d} and In contents \textit{x} determined by XRD of the QWs and QBs for all the samples presented in this work.}
\begin{ruledtabular}
\begin{tabular}{c c c c c c}
sample & substrate & d$_{QW}$ (nm) & x$_{QW}$ & d$_{QB}$ (nm) & x$_{QB}$ \\
\colrule
A&bulk GaN$(0001)$&2.5&0.13&6.3&0.01\\
B&FS-GaN$(000\bar{1})$&3.1&0.23&6.3&0.01\\
C&6H-SiC$(0001)$&2.9&0.14&9.5&0.02\\
D&6H-SiC$(0001)$&5.6&0.14&9.1&0.02\\
E&6H-SiC$(0001)$&8&0.13&9&0.03\\
F&6H-SiC$(0001)$&10.6&0.13&8.9&0.02\\
G&6H-SiC$(000\bar{1})$&2.9&0.19&9.5&0.02\\
H&6H-SiC$(000\bar{1})$&5.8&0.24&9.4&0.01\\
I&6H-SiC$(000\bar{1})$&8.4&0.22&9.4&0.02\\
J&6H-SiC$(000\bar{1})$&11.4&0.20&9.6&0.01\\
\end{tabular}
\end{ruledtabular}
\label{tab:1}
\end{table}

The samples used in this study were independently prepared by two groups (TopGaN and PDI) using different PA-MBE systems and substrates. Samples A and B were prepared at TopGaN. They were grown in a V90 VG Semicon MBE system equipped with a UNI-Bulb Veeco radio-frequency N$_{2}$ plasma source for active N, and solid-source effusion cells for Ga and In. The two samples contain three \ingaxy QWs grown side-by-side on GaN substrates of different polarity. Sample A was grown on a bulk GaN$(0001)$ crystal produced by high nitrogen pressure solution synthesis and sample B on a freestanding GaN$(000\bar{1})$ layer prepared by hydride vapor phase epitaxy. The TD densities of the substrates used for the growth of samples A and B are on the order of $10^{3}$ and $10^{7}$~cm$^{-2}$, respectively. The structural parameters of the QWs and quantum barriers (QBs), as determined by high-resolution x-ray diffraction\cite{Cheze_JVST_01_2013} (HR-XRD), are summarized in Table~\ref{tab:1}. The GaN cap layer is $17$~nm thick and contains less than $1\%$ of In for both samples. The higher In content in the QWs of sample B is the result of the enhanced In incorporation observed for $[000\bar{1}]$-oriented films.\cite{Koblmueller_JAP_01_2007,Nath_01_APL_2010,Cheze_JVST_01_2013} More details about the growth conditions and properties of these samples can be found elsewhere.\cite{Cheze_JVST_01_2013}

Two additional series of samples (C--F and G--J) were grown at PDI in a custom-designed CREATEC PA-MBE system equipped with a UNI-Bulb Veeco radio-frequency plasma source for the generation of active N, and solid-source effusion cells for Ga and In. Samples C--F contain five Ga-polar \ingaxy QWs prepared on 6H-SiC$(0001)$ substrates. The other samples (G--J) contain nominally identical structures on 6H-SiC$(000\bar{1})$, i.\,e., they are of opposite polarity.\cite{Stutzmann_pssb_2001,Wong2013} The typical TD density of GaN layers grown by PA-MBE on 6H-SiC is on the order of $10^{10}$~cm$^{-2}$.\cite{Kaganer_PRB_01_2005,Wong2013} The backside of the substrates was coated with Ti for efficient heat absorption during growth. The as-received SiC substrates were prepared following the procedure described in Ref.~\onlinecite{Brandt1999}. Prior to the growth of the \ingaxy QWs, a $1$~\textmu m thick GaN layer was grown under intermediate Ga-rich growth conditions at $690^{\circ}$C.\cite{Heying_jap_2000} Afterward, the substrate temperature was decreased down to $625^{\circ}$C. At this temperature, we prepared a $100$~nm thick \ingax (x$<0.03$) layer before growing the \ingaxy QWs. The N and In fluxes were kept constant during the uninterrupted growth of QWs and QBs. Their values were $5\times10^{14}$ and $4.3\times10^{14}$~atoms~cm$^{-2}$~s$^{-1}$, respectively. In contrast, the Ga flux was reduced to grow the \ingax QWs, by closing the shutter of one of the two Ga cells used in these experiments, from $4.7\times10^{14}$ to $3.5\times10^{14}$~atoms~cm$^{-2}$~s$^{-1}$. As explained in detail in Ref.~\onlinecite{Siekacz_01_JCG_2008}, this approach facilitates the uninterrupted two-dimensional growth of QWs and QBs under intermediate metal-rich growth conditions.\cite{Gacevic_01_JCG_2013} For all these samples, the nominal width of the barriers and the thickness of the final cap layer were $10.1$ and $30$~nm, respectively. The nominal width of the QWs was varied for each type of substrate between 2.6 and 12~nm. Regardless of the QW width, we observed a streaky reflection high-energy electron diffraction pattern during the growth of the  samples.

The morphological and structural properties of the samples were analyzed by atomic force microscopy (AFM) and HR-XRD, respectively. XRD experiments were performed with Cu$K\alpha_{1}$ radiation (wavelength $\lambda=1.54056$~\AA) using a Panalytical X-Pert Pro MRD system equipped with a Ge$(220)$ hybrid monochromator. Symmetric $\omega$-$2\theta$ scans across the GaN $0002$ Bragg reflection were measured with a three-bounce Ge$(220)$ analyzer crystal. The experimental results were simulated using the dynamical x-ray diffraction model reported in Ref.~\onlinecite{Brandt_JPhysD_2002}.

To gain further insights into the structural properties of the samples, we used TEM. The cross-sectional TEM specimens were prepared by mechanical tripod polishing followed by Ar$^{+}$ milling until reaching electron transparency. All measurements were performed using an aberration-corrected FEI TITAN $80-300$ electron microscope operated at $300$~keV.

The optical properties of the samples were investigated by low-temperature ($10$~K) \textmu-PL and CL spectroscopy. Depending on the well width, PL was excited using either the $413$~nm line of a Kr$^{+}$ laser or the $473$~nm line of a diode-pumped solid-state laser. In both cases, the PL was dispersed by an 80\,cm Horiba Jobin Yvon monochromator and detected by a charge coupled device detector. All measurements were corrected for the spectral response of the PL setup. CL measurements were performed at acceleration voltages of 3--5~kV using a Gatan Mono-CL3 CL system equipped with a parabolic mirror for light collection and with both a photomultiplier and a charge-coupled device (CCD) for detection mounted to a Zeiss Ultra55 field-emission scanning electron microscope (SEM). Monochromatic images were taken with a spectral resolution window of 1--2~nm.

The band profiles and wavefunctions were computed using a self-consistent Schrödinger-Poisson solver.\cite{Snider} To account for the different residual doping of Ga- and N-polar samples,~\cite{Cheze_JVST_01_2013} we set the donor densities in samples A and B to $5\times 10^{16}$ and $1\times 10^{18}$~cm$^{-3}$, respectively. Fermi-level pinning at the surface was taking into account. In accordance with the results reported in Ref.~\onlinecite{Kudrawiec2013}, for our samples exposed to air, we assumed that the Fermi level is pinned $0.6$ and $0.27$~eV below the conduction band minimum for the Ga- and N-polar samples, respectively. Nevertheless, our results do not qualitatively depend on the value at which the Femi level is actually pinned.~\cite{VandeWalle2007,Kudrawiec2013} The InN and GaN band gaps as well as the bowing parameter used to estimate the corresponding values of the \ingax and \ingay layers were extracted from Refs.~\onlinecite{Schley2007} and \onlinecite{King2008}, respectively. For the electron and hole effective masses, we used the values recommended by Vurgaftman and Meyer.\cite{Vurgaftman2003} To calculate the polarization fields, we used the parameters reported in Ref.~\onlinecite{Bernardini2001} by Fiorentini and Bernardini.

\section{Results}

\begin{figure}[t]
\includegraphics[width=0.95\columnwidth]{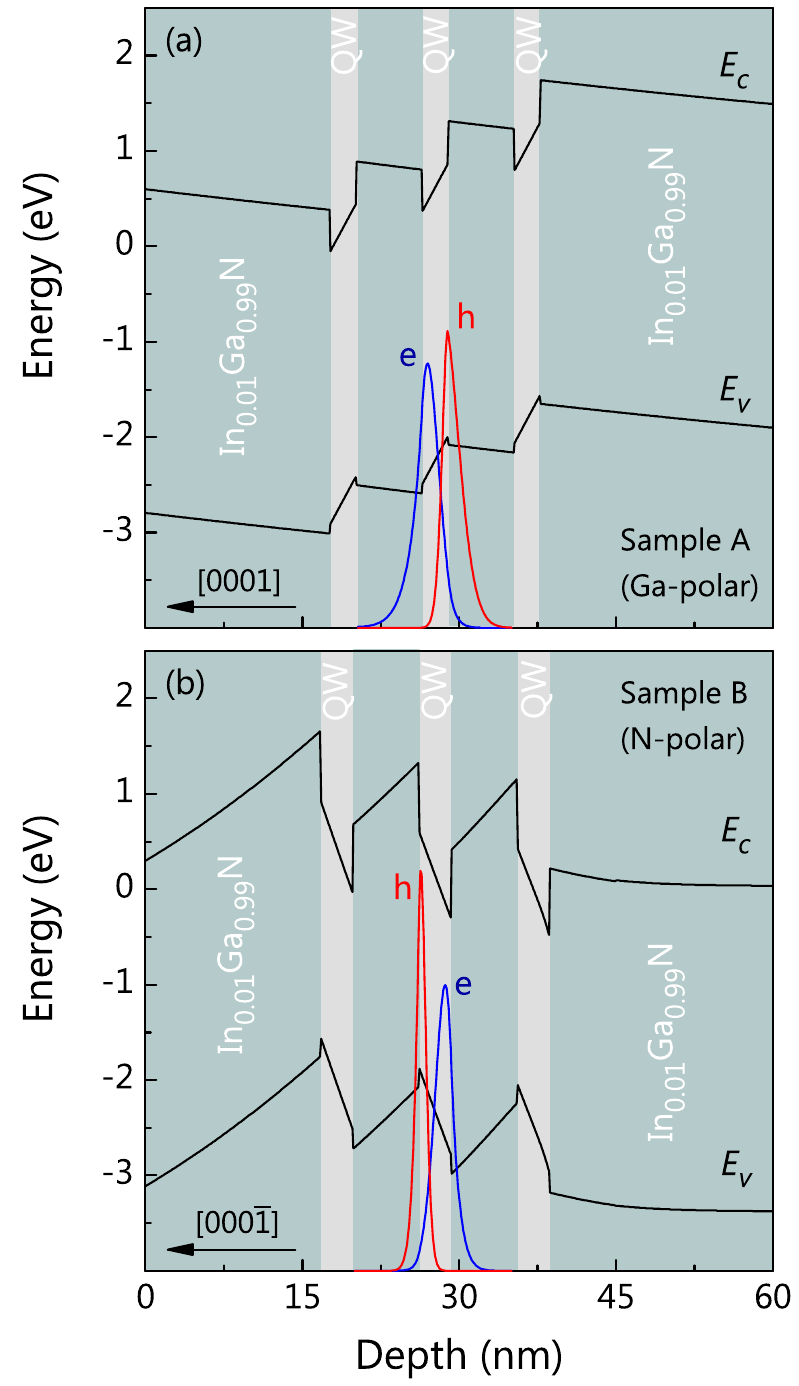}
\caption{\label{bands}(Color online) Band profiles of the Ga-polar sample A (a) and the N-polar sample B (b). In both figures, we also show the calculated electron (solid blue line) and hole (solid red line) wave functions for the second QW.}
\end{figure}

\subsection{Carrier escape and surface recombination as possible origin of the lack of PL in N-polar \ingaxy QWs}

To investigate whether the lack of PL for N-polar \ingaxy QWs reported by \citet{Cheze_JVST_01_2013} may be caused by carrier escape from the QWs, we calculated the band profiles of Ga- and N-polar QWs, and performed low-temperature ($10$~K)~\textmu-PL experiments with resonant excitation.

\subsubsection{Simulation of the band profiles}

Figure~\ref{bands} presents the simulated band profiles for the Ga- and N-polar QWs studied by \citet{Cheze_JVST_01_2013} (samples A and B). We also present the calculated ground-state electron and hole wave functions for the second QW. Similar results were obtained for the other two QWs (not shown here for clarity).

For the N-polar \ingaxy QWs (sample B), the electric field near the sample surface is stronger and of opposite direction compared to their Ga-polar counterparts. In the cap layer, electrons are thus pulled toward the surface and holes dragged into the QW region. Due to the reverse direction of the piezoelectric fields, the direction of the electric field inside the QWs is also opposite for samples A and B. Because of this reason, electron and hole wave functions are localized at opposite QW/QB interfaces.

The simulations demonstrate that the electron and hole ground states are well confined into the QWs for both samples. To reach the surface, both electrons and holes would have to overcome large barriers, particularly so for sample B. In fact, comparing the spatial extension of the wavefunctions, the confinement of both electrons and holes is seen to be significantly stronger in the QWs of sample B as a result of the reversed fields and the higher In content (see Table~\ref{tab:1}). An escape of carriers from the QWs seems hardly possible for sample A, but much less so for sample B.

Finally, we note that the overlap between the electron and hole wave functions is about a factor of two larger for sample A as compared to sample B. This reduction in overlap, due to the slightly larger QW thickness as well as the stronger electric field caused by the higher In content, could explain a corresponding quenching of the luminous efficiency (assuming an equally strong nonradiative participation in recombination), but certainly not the absolute lack of luminescence reported by \citet{Cheze_JVST_01_2013}

\subsubsection{Low-temperature photoluminescence with resonant excitation}
\label{ltpre}

\begin{figure}[t]
\includegraphics[width=0.9\columnwidth]{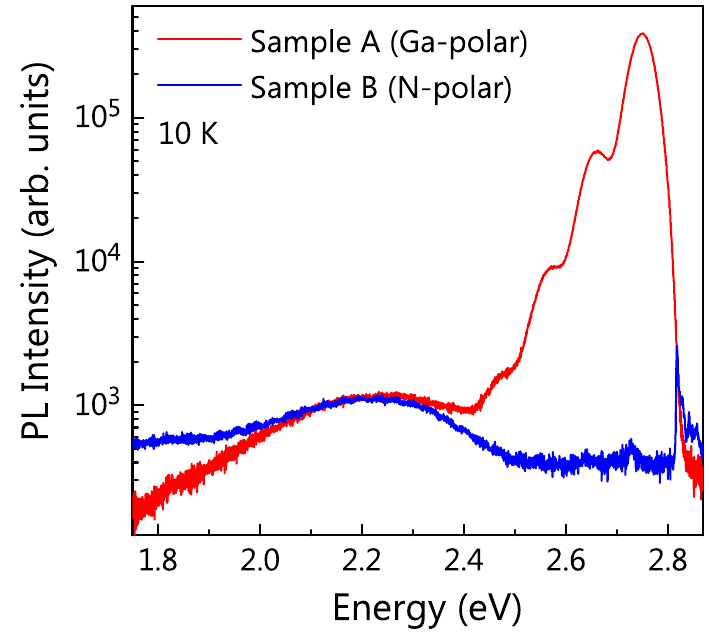}
\caption{\label{PL1}(Color online) Low-temperature (10~K) \textmu-PL spectra under resonant excitation (413~nm) of samples A and B. The samples were measured side-by-side with an excitation density of 200 and 2000~kW/cm$^2$ for samples A and B, respectively. The sharp feature at 2.82\,eV for sample B originates from second order optical phonon scattering in the GaN substrate. The spectra have been vertically shifted to line up their background in order to account for the different integration times used during the measurements.}
\end{figure}

The low-temperature (10~K) PL experiments reported by \citet{Cheze_JVST_01_2013} were performed using the $325$~nm line of a He-Cd laser. For this wavelength, most carriers are not directly excited in the QWs but in the cap layer and the surrounding QBs. The QW emission may then be suppressed by nonradiative recombination in the QB, competing with the capture of carriers by the QWs. To exclude this possibility, we here examine the PL spectra of samples A and B obtained by \emph{direct} excitation at $413$~nm.

Figure~\ref{PL1} shows the low-temperature PL spectra of samples A and B recorded with direct excitation. For sample A, we observe a strong emission at $\approx2.75$~eV as well as its longitudinal optical (LO) phonon replica from the Ga-polar \ingax QW. The yellow band centered at $\approx2.25$~eV stems from the GaN substrate. For sample B, we observe only the yellow band at $\approx2.25$~eV and a sharp feature at 2.82\,eV which originates from second order optical phonon scattering but no emission line which could be associated to the N-polar \ingax QW.

\subsection{Effect of quantum well width on the properties of Ga- and N-polar \ingaxy QWs}

The results presented in the previous section demonstrate the presence of an efficient nonradiative recombination channel within N-polar \ingaxy QWs. Because \citet{Cheze_JVST_01_2013} observed an intense luminescence from thick N-polar \ingax layers grown under similar conditions (comparable to the intensity observed in Ga-polar \ingaxy QWs), the most plausible explanation for the lack of luminescence is a high concentration of nonradiative defects at the QW/QB interfaces. Within the classical treatment of surface and interface recombination,\cite{Ahrenkiel_1993} and assuming a diffusion length much larger than the layer thickness, the nonradiative rate is expected to decrease linearly with the thickness of the layer. In the present case, the location of the charge carriers is determined not by diffusive transport but by drift in the strong internal electrostatic field within the \ingax QWs, and it is thus doubtful whether this expectation still holds. To investigate this scenario, we analyze the luminescence of nominally identical Ga- and N-polar \ingaxy structures as a function of the width of the \ingax QW (samples C--J).

\subsubsection{Morphological and structural characterization}
\label{msc}

\begin{figure}[t]
\includegraphics[width=0.8\columnwidth]{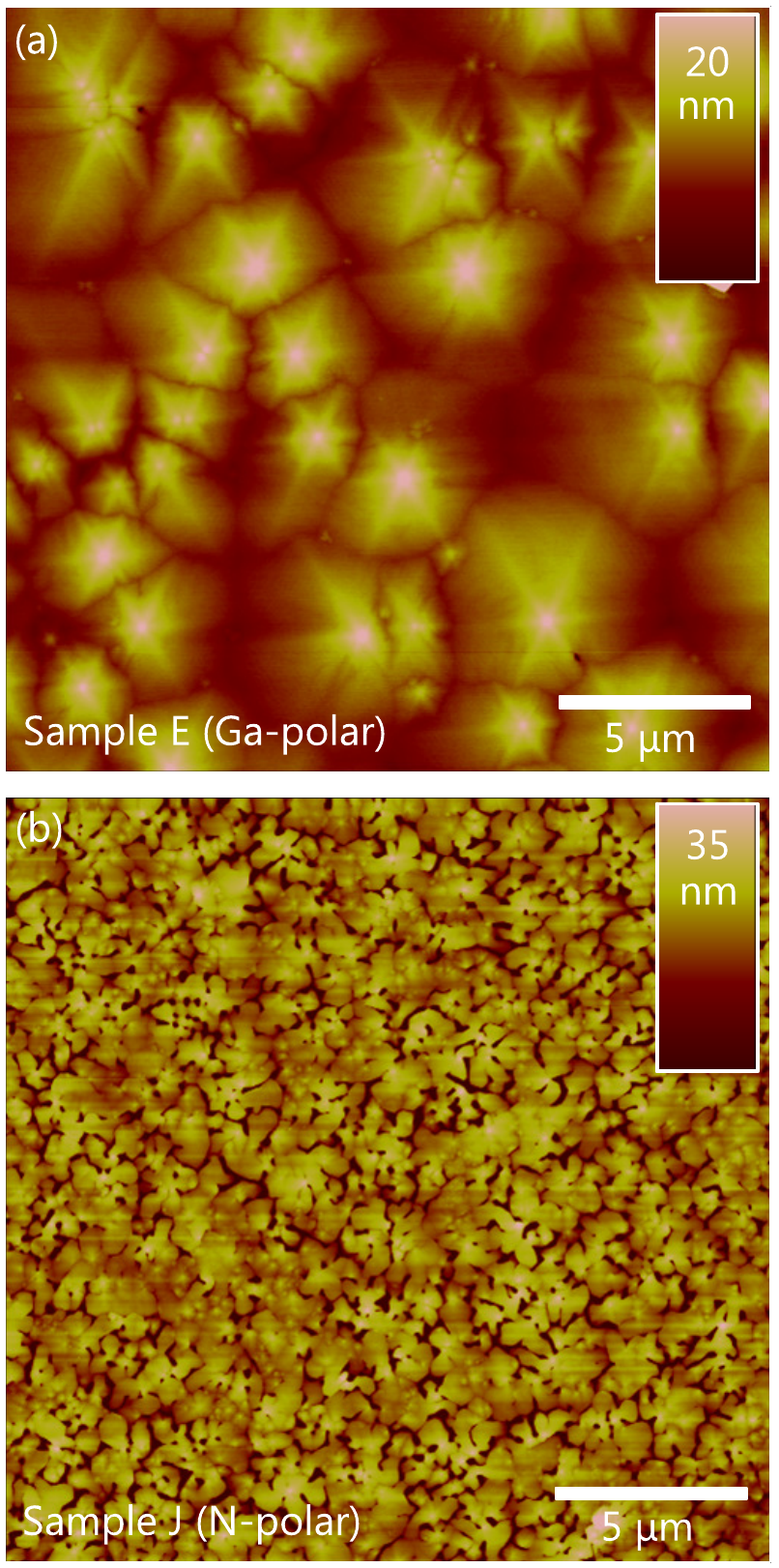}
\caption{\label{AFM}(Color online) Characteristic AFM images of Ga- (a) and N-polar (b) \ingaxy QWs on SiC\{0001\}. The images shown were taken from samples E (a) and J (b).}
\end{figure}

\begin{figure}[t]
\includegraphics[width=0.95\columnwidth]{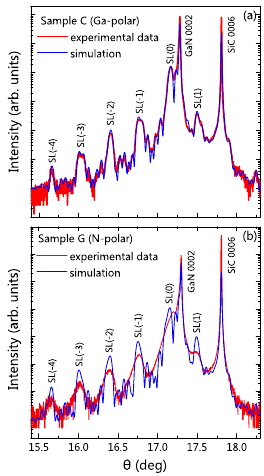}
\caption{\label{XRD} (Color online) Examples for  experimental and simulated triple-crystal $\omega$-$2\theta$ scans across the GaN 0002 reflection of Ga- (a) and N-polar (b) \ingaxy QWs on SiC\{0001\}. The scans shown were taken from samples C (a) and G (b).}
\end{figure}

The morphological and structural properties of the QW samples prepared on SiC were investigated by AFM and XRD. Figures~\ref{AFM}(a) and \ref{AFM}(b) show characteristic $20\times20$~\textmu m$^{2}$ AFM images of two QW samples grown along the $[0001]$ and $[000\bar{1}]$ directions, respectively. For the Ga-polar QWs (samples C--F), the surfaces are characterized by atomic steps and hillocks. The latter originate from spiral step-flow growth around screw dislocations. The density of hillocks is on the order of $\approx1\times10^{7}$~cm$^{-2}$ and the root-mean-square roughness (rms) is lower than 1~nm. The surfaces of the N-polar QWs (samples G--J) are much rougher with rms values as high as 8~nm. These larger values are primarily due to the presence of a high density of pits as shown in Fig.~\ref{AFM}(b).

Figure~\ref{XRD} presents two examplary $\omega$-$2\theta$ scans performed to assess the In concentration as well as the widths of the QWs and QBs of samples C--J. The figure also shows the corresponding simulated XRD profiles. The parameters derived from the simulations, assuming that QWs and QBs have the same in-plane lattice constant as the GaN layer underneath, are summarized in Table~\ref{tab:1}. For the Ga-polar \ingaxy QWs [Fig.~\ref{XRD}(a)], we observe satellite peaks up to at least fourth order in addition to the dominant SiC $0006$ and GaN $0002$ reflections. The excellent agreement of experimental and simulated profiles, including the complex interference fringes modulating the satellites, reflects the high periodicity and the abrupt interfaces of the \ingaxy quintuple QW structure. The structural parameters for the Ga-polar samples C--F are thus obtained with high accuracy. For the N-polar \ingaxy QWs, we also observe satellite peaks up to fourth order [see Fig.~\ref{XRD}(b)], but all satellites (including the zeroth order one) are significantly broadened compared to the simulation. This broadening indicates a significant compositional variation within the structure. Due to this broadening, the error in the structural parameters obtained for the N-polar samples G--J is larger than for their Ga-polar counterparts. However, we consistently obtain QW and QB widths comparable to those of samples C--F, and a higher In content $x$ [on average $(21.3 \pm 2.2) $\% as compared to $(13.5 \pm 0.6)$\% for samples C--F]. This latter result reflects again the enhanced In incorporation efficiency along the $[000\bar{1}]$ direction as reported previously by several groups.\cite{Koblmueller_JAP_01_2007,Nath_01_APL_2010,Cheze_JVST_01_2013} 

\begin{figure*}
\centering
\includegraphics[width=15.5cm]{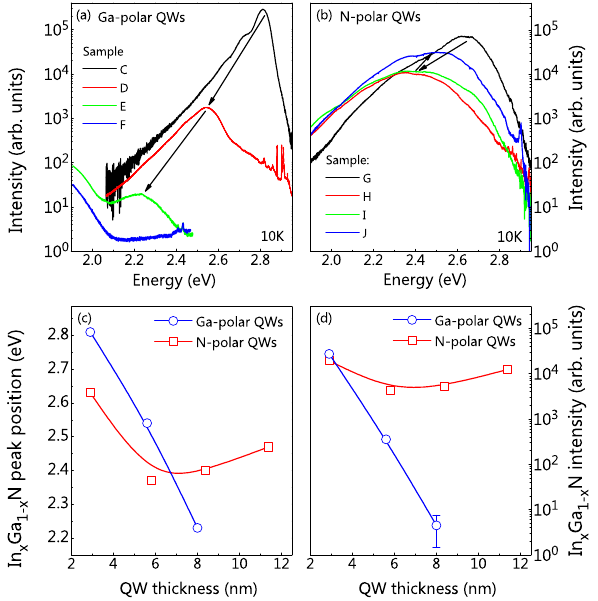}
\caption{\label{PL} (Color online) Low-temperature (10~K) \textmu-PL spectra of Ga- (a) and N-polar (b) \ingaxy QWs with different width. The arrows indicate the shift of the \ingax peak with increasing QW width. The energy position  and integrated intensity of the QW-related emission in the \textmu-PL spectra is shown in (c) and (d), respectively, as a function of the QW width.}
\end{figure*}

\subsubsection{Low-temperature photoluminescence with resonant excitation}

Figure~\ref{PL} presents the low-temperature (10~K) \textmu-PL spectra excited at 413~nm (473~nm for samples E and F) of the Ga- (a) and N-polar (b) \ingaxy QWs. For the Ga-polar samples [Fig.~\ref{PL}(a)], a strong QW-related emission centered at $2.81$ and $2.54$~eV was observed for the two samples with the narrowest QWs, i.\,e., sample C and sample D, respectively. For sample E, the donor-acceptor-pair (N-Al) emission of SiC dominated the spectrum for excitation at 413~nm (not shown). When exciting with the 473~nm line, we detect a weak but clear emission from the \ingax QWs centered at 2.23~eV. Finally, for sample D, we did not observe any emission that could be ascribed to the QWs. The luminescence band below 2~eV, which is also seen in the PL spectra of the other samples, is due to partial dislocations in faulted regions in the SiC substrate.\cite{Galeckas2006} The N-polar samples, the spectra of which are shown in Fig.~\ref{PL}(b), behave completely different: for all samples, we detect a broad emission band much more intense than any emission from SiC. Neither the energy position nor the intensity of this band seems to depend systematically on QW width.

\begin{figure}[t]
\includegraphics[width=0.65\columnwidth]{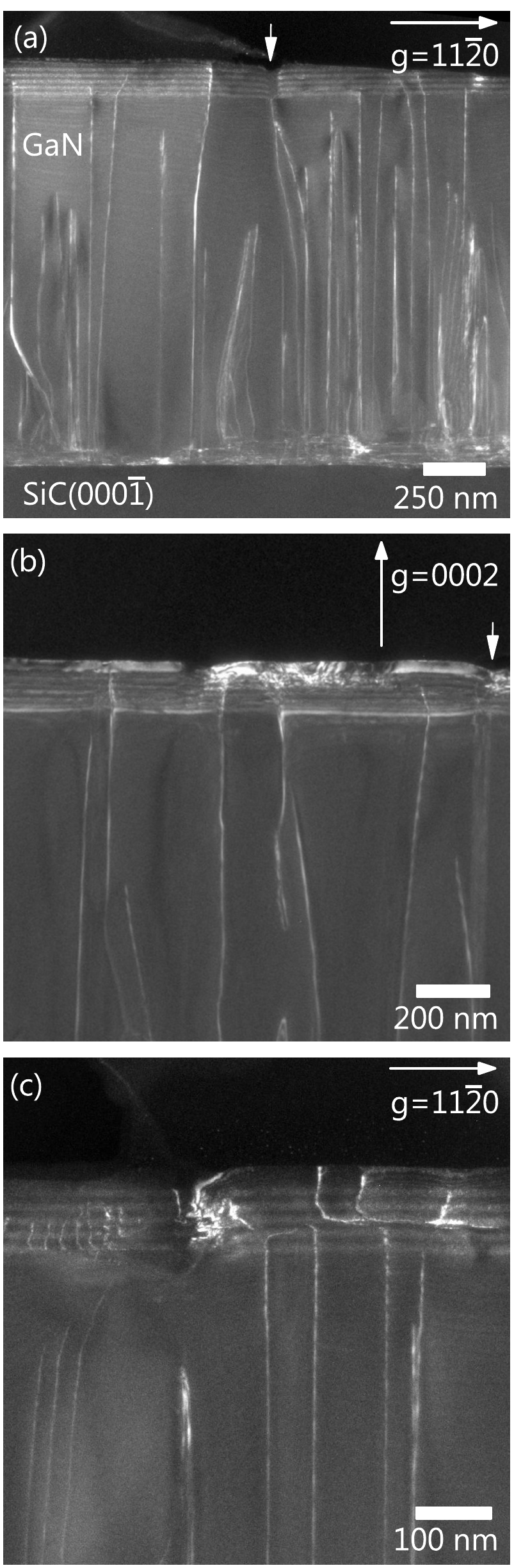}
\caption{\label{TEM1}(Color online) WBDF micrographs of sample I showing the entire sample structure (a) and only the PA-MBE grown layers (b) and (c). The diffraction conditions are indicated in each figure. Mixed dislocations are visible in all of the images. \textit{a}-type TDs are visible only in (a) and (c); \textit{c}-type TDs are visible only in (b). The small arrows indicate the position of the $\vee$-pit defects.}
\end{figure}

\begin{figure}[t]
\includegraphics[width=0.75\columnwidth]{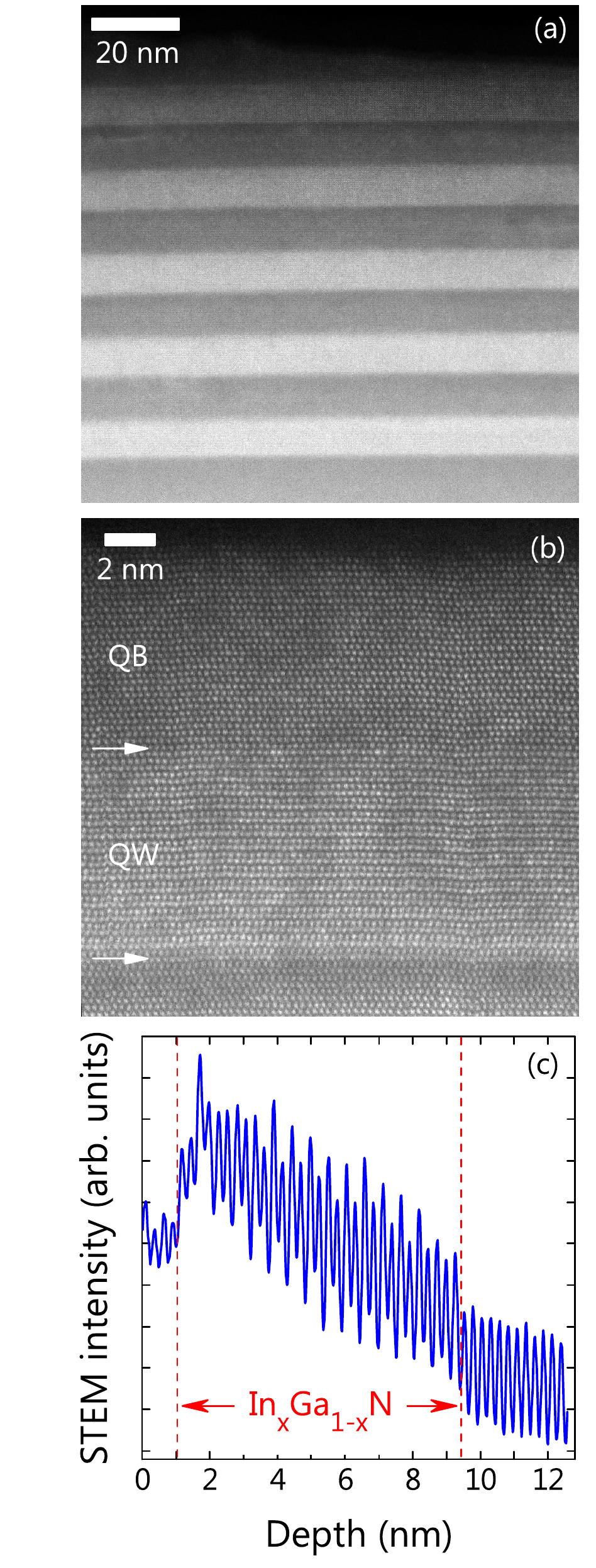}
\caption{\label{TEM2}(Color online) STEM-HAADF micrographs of the entire QW region (a) and the fifth QW (b) of sample I. In (b), the arrows indicate the position of the QW/QB interface. Both micrographs were taken along the $[11\bar{2}0]$ direction. (c) STEM-HAADF intensity profile along the growth direction for the fifth QW. The red dashed lines indicate the QW/QB interfaces.}
\end{figure}
\begin{figure}[t]
\includegraphics[width=\columnwidth]{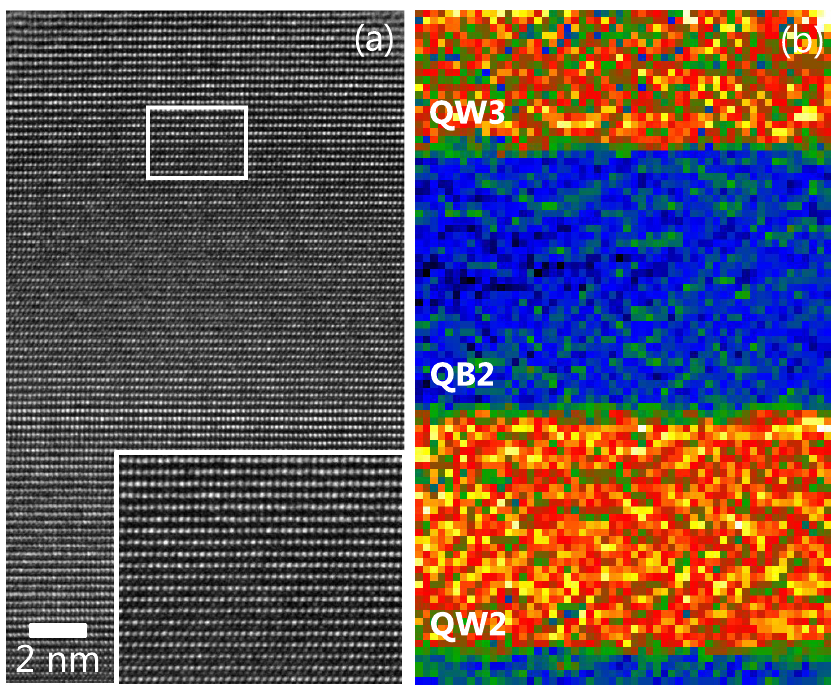}
\caption{\label{TEM3}(Color online) (a) HRTEM image of the QW region of sample I taken along the $[1\bar{1}00]$ direction. The inset shows with higher magnification the interface between the third QW and the subsequent QB. (b) Color coded map of the local lattice parameter \textit{c} distribution ranging from 500 pm (blue) to 550 pm (yellow).}
\end{figure}

\begin{figure}[t]
\includegraphics[width=0.95\columnwidth]{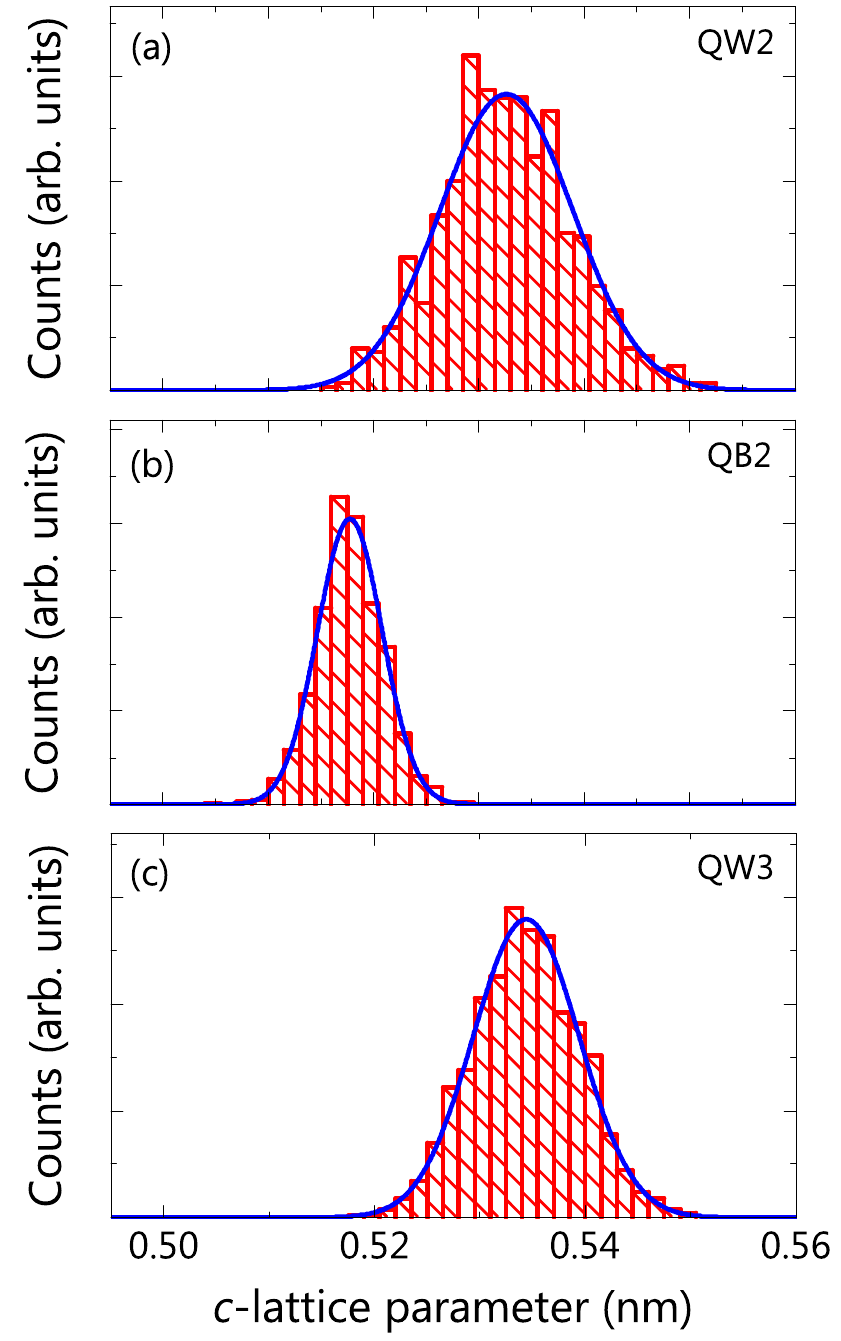}
\caption{\label{TEM4}(Color online) Histograms of the local lattice parameter \textit{c} measured in a QB and two different QWs of sample I. The solid blue lines represent Gaussian fits to the experimental data.}
\end{figure}

\begin{figure}[t]
\includegraphics[width=0.75\columnwidth]{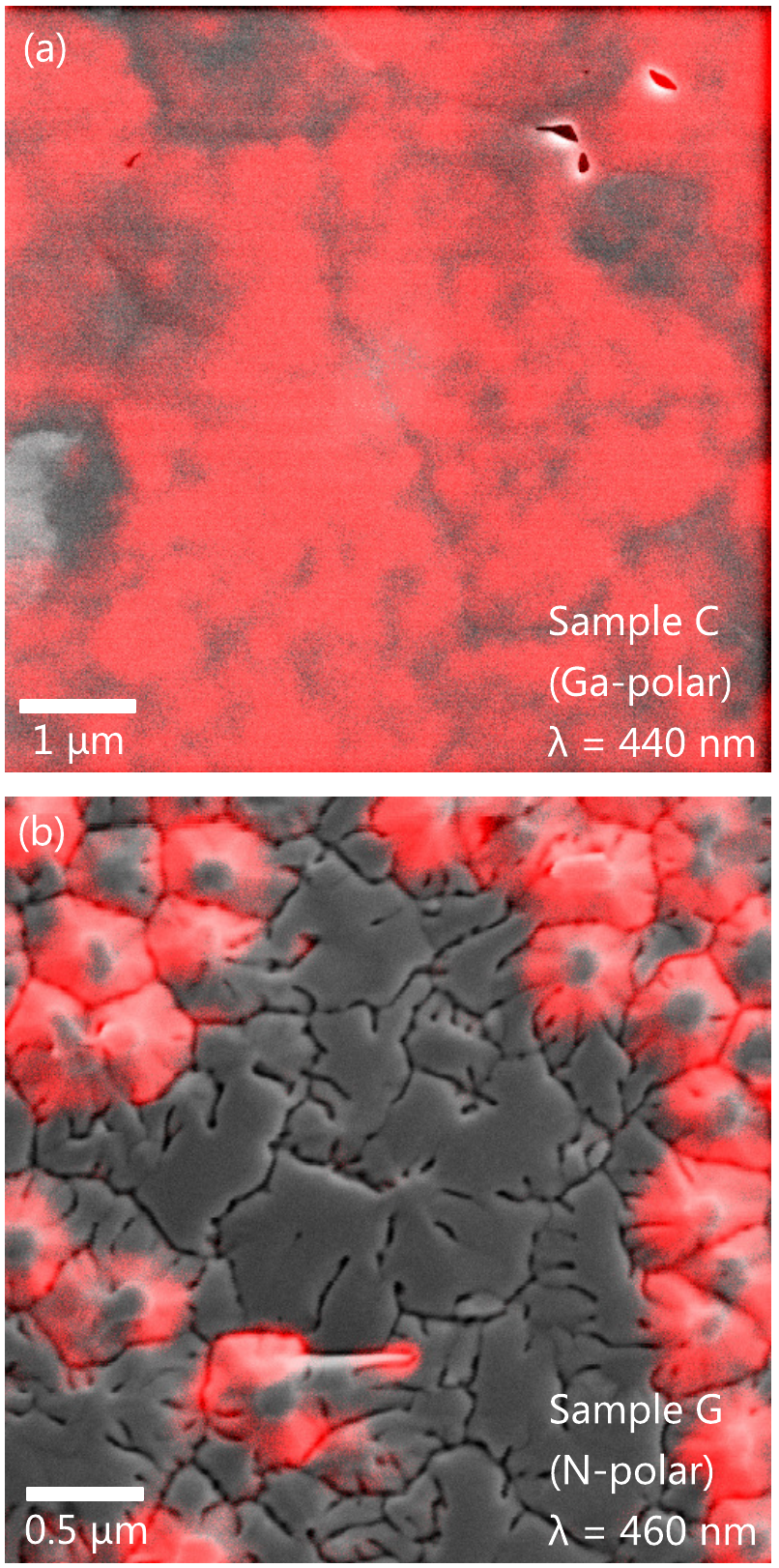}
\caption{\label{CL}(Color online) Representative top-view monochromatic CL and superimposed SEM images of Ga- (a) and N-polar (b) \ingaxy QWs. The images shown are taken from samples C (a) and G (b). The CL images were taken at 10~K. For the N-polar sample G, the luminescence observed stems exclusively from the semi-polar facets formed around TDs.}
\end{figure}

Figures~\ref{PL}(c) and \ref{PL}(d) show the energy and the integrated intensity of the \ingax emission as a function of the QW width for the two series of samples. For the Ga-polar samples, both quantities exhibit the expected trend, namely, the peak red-shifts and quenches with increasing width due to the quantum-confined Stark effect (QCSE). In contrast, for the N-polar samples, the peak does neither exhibit a continuous red-shift nor quench with increasing QW width. In fact, the N-polar samples are brighter than their Ga-polar counterparts with the sole exception of the sample with the thinnest QWs.

These results are in striking contrast to those presented in Sec.~\ref{ltpre} and those reported in Ref.~\onlinecite{Cheze_JVST_01_2013}. The independence of PL energy and intensity on QW width observed for samples G--J could be understood if the emission would be dominated by strongly localized states created by non-random alloy  inhomogeneities.\cite{Chichibu1996,Narukawa1997,SeoIm1998,Chichibu1997,ODonnell1999,Chichibu2006} Strong alloy fluctuations would also be consistent with the broadening of the satellite peaks observed in the triple-crystal $\omega$-$2\theta$ scans discussed in Sec.~\ref{msc}. In the following, we therefore investigate the samples at a microscopic and nanometric scale by TEM. In addition, we attempt to unravel the microscopic origin of the emission observed by spatially-resolved CL spectroscopy.

\subsubsection{Transmission electron microscopy}

To study the microscopic structure of the N-polar \ingaxy QWs, we analyzed samples G and I by various TEM techniques. Weak-beam dark-field (WBDF) imaging was used to visualize the spatial distribution of different types of dislocations as well as to estimate TD densities. The atomic structure of the QWs was studied by high-angle annular dark-field scanning transmission electron microscopy (STEM-HAADF). High-resolution TEM (HRTEM) was used for a structural analysis of the alloys on an atomic scale.

Figures~\ref{TEM1}(a)--\ref{TEM1}(c) show cross-sectional WBDF micrographs taken of sample I with different diffraction vectors. The micrographs show that some but not all of the TDs originating at the GaN/SiC interface affect the growth of the \ingaxy QWs and lead to the formation of $\vee$-pit defects also visible in the AFM images depicted in Fig.~\ref{AFM}(b). The comparison of Figs.~\ref{TEM1}(a) and \ref{TEM1}(b) reveals that the ratio between \textit{c}- and \textit{a}-type TDs is approximately 1:5, while their total areal density is in the range of 1--4$\times 10^{10}$~cm$^{-2}$. This value is comparable to the typical TD densities reported in the literature for GaN layers grown on 6H-SiC.\cite{Kaganer_PRB_01_2005,Wong2013} Quite unexpectedly, Fig.~\ref{TEM1}(c) also reveals the presence of \emph{a}-type misfit dislocations (MDs) in sample I, well below the onset of significant plastic relaxation of the entire film in \ingax layers on GaN.\cite{Amano_1996,Parker_APL_1999,Pereira_APL_2002,Dobrovolskas_JAP_2013} Note, however, that plastic relaxation is only observed locally. Local plastic relaxation can be induced by spatial variations in the growth mode caused by fluctuations in the growth conditions or by the presence of structural and/or morphological defects. Nevertheless, the majority of the interfacial area remains coherently strained as further discussed below. Also note that we did not observe any MDs for sample G (not shown here).

Figure~\ref{TEM2} shows STEM-HAADF images of the QW region of sample I taken in a region free from $\vee$-pits and MDs, where the QWs exhibit a planar morphology. High-resolution STEM-HAADF images [Fig.~\ref{TEM2}(b)] do not show any indication for fluctuations in the In content along the basal (0002) plane. The intensity profile along the growth direction displayed in Fig.~\ref{TEM2}(c) shows that both the lower and the upper QW/QB interfaces are abrupt on an atomic scale (the linear change in the STEM intensity along the growth direction is due to the continuous variation in the specimen width). The broadening of the satellite peaks observed in the XRD profiles [Fig.~\ref{XRD}(b)] is thus not related to microscopic interface roughness. As shown in Fig.~\ref{TEM2}(b), the QW contains 31 atomic planes of \ingax. The widths of the QWs from samples G and I thus measured by STEM-HAADF (3 and 8.4~nm, respectively) are in good agreement with the results derived from the simulations of the corresponding XRD profiles (see Table~\ref{tab:1}).

We analyzed the same sample region by HRTEM in order to further examine the crystal quality of the active region, and in particular to check the compositional fluctuations of the \ingaxy QWs. Figures~\ref{TEM3}(a) presents a HRTEM image of the second and third QWs from sample I. A Fourier filtered image (not shown here) confirms the absence of extended defects, such as misfit dislocations or stacking faults, within this region.

Figure~\ref{TEM3}(b) shows the distribution of the local lattice parameter \textit{c} extracted from a series of HRTEM images similar to the one shown on the Fig.~\ref{TEM3}(a). The local lattice parameter was obtained using the algorithm developed in Ref.~\onlinecite{Schulz_01_JAP_2012}. Since the QWs are coherently strained, the local lattice parameter \textit{c} reflects the distribution of the In content inside the QWs. In order to check the compositional uniformity, we construct histograms of the lattice parameter \textit{c} for two QWs and one QB from sample I as shown in Fig.~\ref{TEM4}. For both QWs and the QB, the histograms are well described by Gaussian distributions indicating that the In distribution in our N-polar \ingaxy QWs is essentially random.\cite{Schulz_01_JAP_2012} Hence, the anomalous dependence of emission energy and intensity cannot be ascribed to carrier localization by strong, non-random alloy fluctuations. 

\subsubsection{Cathodoluminescence}

To elucidate the microscopic origin of the luminescence, we examined the samples simultaneously by SEM and low-temperature (10~K) CL. Figures~\ref{CL}(a) and \ref{CL}(b) show, for comparison, the superposition of the scanning electron micrographs with the monochromatic CL maps for samples C and G, respectively, for representative spectral windows centered around the peak of the \ingax emission band. For the Ga-polar sample C [Fig.~\ref{CL}(a)], the surface is flat and smooth, in good agreement with the AFM measurements. The \ingax-related CL signal is distributed rather uniformly over the entire surface. 

The surface of the N-polar sample G [Fig.~\ref{CL}(b)] consists of flat and smooth $(000\bar{1})$ facets separated by narrow trenches. In addition, a high density of $\vee$-pits are visible which are seen to have the shape of an inverted pyramid with semipolar facets. The superposition of the scanning electron micrograph with the monochromatic CL map reveals that the \ingax-related luminescence does not originate from the flat and smooth $(000\bar{1})$ facets but from the semi-polar facets that constitute the $\vee$-pit defects. In fact, also in monochromatic imaging using other spectral windows or in spectrally resolved measurements on the $(000\bar{1})$ facets (not shown here), we did not observe any CL signal that could be ascribed to the N-polar \ingaxy QWs.

The reduced piezo-electric field along semipolar directions\cite{Romanov2006} partly explains the nonsystematic variation of the PL spectrum when increasing the QW width from sample G to J. In any case, the understanding of this variation is, at this point, only of secondary importance. The important result here is the lack of luminescence from the flat $(000\bar{1})$ facets that correspond to the regions with well-defined N-polar \ingaxy QWs. This observation confirms the presence of a highly efficient nonradiative recombination channel within the flat regions of the QWs.

\section{Discussion}

If N-polar \ingaxy QWs are to be used in light-emitting devices, the origin of this nonradiative channel has to be identified and eliminated. Evidently, TD cannot be at the root of this nonradiative channel since the N-polar \ingaxy QWs studied in Ref.~\onlinecite{Cheze_JVST_01_2013} have a TD density of only $10^7$~cm$^{-2}$, three orders of magnitude lower than samples G--J which were grown on SiC substrates. Yet, as seen in Fig.~\ref{PL1}, these samples were also found to exhibit no \ingax-related emission at all. Misfit dislocations cannot explain our results either because they were not observed in sample G. A higher concentration of nonradiative recombination centers in the bulk of the film, associated to the more efficient incorporation of impurities on the $(000\bar{1})$ plane, is unlikely because \citet{Cheze_JVST_01_2013} observed an intense luminescence from thick N-polar \ingax layers grown under similar conditions as those used for QWs. We thus believe that the most plausible explanation left is a high concentration of nonradiative point defects located at the QW/QB interfaces. In fact, it is known for other III-V compound semiconductor heterostructures \cite{Behrend1996} that point defects can segregate on the growth front, and may incorporate at the heterointerface due to the abrupt change of electrochemical potential. This mechanism is particular to a given surface, and may result in very high point defect densities at either the normal or the inverted interface between two materials. A high density of nonradiative point defects at either the \ingaxyo or the \ingayxo interface would be compatible with all experimental results reported in Ref.~\onlinecite{Cheze_JVST_01_2013} and the present work.

Our results and those reported in Ref.~\onlinecite{Cheze_JVST_01_2013} contrast with the successful fabrication of N-polar \ingax LEDs by \citet{Akyol_01_JJAP_2011} However, these authors did not analyze the microscopic origin of the electroluminescence from their devices. The emission may thus have originated from semipolar QWs formed around $\vee$-pits as observed in the present work or around hillocks as reported in Ref.~\onlinecite{Song_01_AMI_2015}.

The unexpected luminescence quenching observed in N-polar \ingaxy QWs is not only relevant for the fabrication of LEDs in the form of films but is also of importance for the development of nano-LEDs based on spontaneously formed GaN nanowires (NWs) in PA-MBE. As discussed in Refs.~\onlinecite{Hestroffer_2011,Fernandez-Garrido_nl_2012,Romanyuk2015}, these nanostructures, if formed spontaneously in the absence of structural or morphological defects of the substrate, elongate along the $[000\bar{1}]$ direction. GaN NWs are thus N-polar and so are \ingax quantum disks inserted into them. Yet, numerous publications report on the luminescence of these N-polar \ingax disks embedded into GaN nanowires.\cite{Kikuchi2004a,Tourbot2011,Woelz2012,Tourbot2012,Laehnemann2014} While the luminous efficiency of these structures does not seem to be even close to that of planar Ga-polar \ingax QWs,\cite{Tourbot2011,Woelz2012,Tourbot2012,Marquardt_01_NL_2013,Laehnemann2014} they do luminesce in contrast to the planar QWs studied in the present work. Two essential differences between the \ingax disks inserted in GaN NWs and corresponding planar structures may contribute to this fact: first, GaN NWs are formed under N excess,\cite{Fernandez-Garrido2009,Fernandez-Garrido_NL_2013} while planar QWs are grown under metal-stable conditions. The resultingly different stoichiometry at the growth front may very well suppress the formation of nonradiative defects during the growth of the \ingax quantum discs. Second, \ingax disks in GaN NWs have been shown to exhibit large compositional fluctuations,\cite{Tourbot2011,Laehnemann_PRB_2011,Tourbot2012} inducing carrier localization which in turn may at least partly prevent nonradiative recombination.

Last but not least, we note that the conditions employed here and in Ref.~\onlinecite{Cheze_JVST_01_2013} for the growth  of N-polar \ingaxy QWs are those considered to be optimal for their Ga-polar counterparts. These conditions result in an In-terminated surface which is essential for establishing a two-dimensional growth front on the $(0001)$ surface, but may very well result in an enhanced generation of point defects on the $(000\bar{1})$ surface. Clearly, further studies are needed to investigate the impact of the surface stoichiometry during growth on the luminous efficiency of N-polar \ingaxy QWs. A promising route to improve the quality of N-polar QWs consists in taking advantage of their higher thermal stability for exploring growth at substrate temperatures not achievable in the growth of Ga-polar \ingax.\cite{cheze_privatecomm}

\section{Summary and conclusions}

We have analyzed the properties of Ga- and N-polar \ingaxy QWs grown by PA-MBE under nominally identical conditions on different types of substrates. Ga-polar QWs exhibit a strong luminescence. Both the energy and the intensity of the luminescence are largely determined by the QCSE, and thus by the width of the QWs. In contrast, regardless of the substrate as well as of the QW width, we did not observe luminescence from N-polar QWs. The simulation of the band profiles together with PL measurements carried out with resonant excitation point toward the presence of a very efficient nonradiative recombination channel within the QWs. Because the luminescence quenching is not observed in thick N-polar \ingax layers, the most plausible explanation for this phenomenon is the incorporation of a high density of nonradiative defects at the interfaces between QWs and QBs. The present results demonstrate that the optimized growth conditions for the fabrication of Ga-polar \ingaxy QWs are not suitable for the synthesis of equivalent structures along the opposite direction. In order to take advantage of the potential benefits of the $[000\bar{1}]$ orientation for the fabrication of green LEDs, it is necessary to develop novel growth approaches designed to eliminate the nonradiative defects in N-polar \ingaxy QWs.

\section{Acknowledgments}

We would like to thank Hans-Peter Schönherr for his dedicated maintenance of the MBE system, and Henning Riechert for a critical reading of the manuscript. Partial financial support from the European Union through grant SINOPLE 230765 is gratefully acknowledged.

\end{document}